\documentstyle[floats,twocolumn,aps,prl,psfig,pstricks] {revtex}

\begin{document}

\title{{\it Ab initio} molecular dynamics study of the desorption 
of D$_2$ from Si\,(100)}

\author{Axel Gross, Michel Bockstedte, and Matthias Scheffler}

\address{Fritz-Haber-Institut der Max-Planck-Gesellschaft, Faradayweg 4-6,
D-14195~Berlin-Dahlem, Germany}

\twocolumn[

\maketitle

\vspace{-.7cm}

\begin{abstract}

\hspace*{1.5cm}\parbox{15cm}{
{\it Ab initio} molecular dynamics calculations 
of deuterium desorbing from Si\,(100) have been performed in order to 
monitor the energy redistribution among the various D$_2$ and silicon 
degrees of freedom during the desorption process. The calculations 
show that a considerable part of the potential energy at the transition
state to desorption is transferred to the silicon lattice.
The deuterium molecules leave the surface vibrationally hot and
rotationally cold, in agreement with thermal desorption experiments; 
the mean kinetic energy, however, is larger than found in 
a laser-induced desorption experiment. We discuss possible
reasons for this discrepancy.
}
\end{abstract}

\hspace*{1.5cm}\parbox{15cm}{
}

\vspace{-1.0cm}

]

Hydrogen adsorption on and desorption from Si surfaces are of great
technological relevance for, e.g., the etching and passivation of
Si surfaces or the growth of Si crystals
(see, e.g., Ref.~\cite{Kol96} and references therein). 
Besides, the dynamics of the hydrogen interaction with Si surfaces is 
also of fundamental interest caused, among others, 
by the so-called barrier puzzle:
While the sticking coefficient of molecular hydrogen on Si surfaces
is very small \cite{Lie90,Kol94b,Bra95,Bra96a}
indicating a high barrier to adsorption, in desorption experiments
an almost thermal mean kinetic energy  of the molecules was found 
\cite{Kol94a} indicating a low adsorption barrier. In order to 
explain this puzzle it was suggested to take the strong surface rearrangement 
of Si upon hydrogen adsorption into account \cite{Kol94a}: The hydrogen 
molecules impinging on the Si substrate from the gas phase typically
encounter a Si configuration which is unfavorable for dissociation, 
while desorbing hydrogen molecules leave the surface from a rearranged 
Si configuration with a low barrier.
Unfortunately, at this point it is unclear how low the barrier really is,
because the results for the dissociative adsorption probability of references
\cite{Kol94b} and  \cite{Bra96a} differ by almost three orders of magnitude.
This difference translates into a difference of the minimum barrier heights
measured in these two experiments of about 0.4~eV.

Total-energy calculations using the cluster approach 
\cite{Wu93,Nac94,Jin95,Rad96} have found activation barriers for 
associative desorption of about 1~eV. 
With the assumption of a lattice rearrangement energy of about 0.7~eV 
the experimental adsorption results of Refs.~\cite{Bra95,Bra96a} {\em and} 
the desorption results of Ref.~\cite{Kol94a} could be 
reproduced by quantum dynamical model calculations \cite{Bre94,Bra96}.
Still the exact adsorption and desorption mechanism is strongly debated 
in the system H$_2$/Si\,(100). In view of density-functional theory (DFT)
calculations for Si\,(100) a lattice-relaxation energy of 0.7~eV seemed
to be too high \cite{Dab92}.
Indeed, in detailed DFT  calculations of the H$_2$/Si\,(100)
potential energy surface (PES) using the supercell 
approach~\cite{Kra94,Peh95,Vit95} the adsorption barriers were found
to be only 0.3~-~0.4~eV with a substrate rearrangement energy 
of about 0.15~eV.

Based on the slab calculations \cite{Kra94}
and approximating the high-dimensional PES by a three-dimensional one,
two different quantum dynamical studies were performed \cite{Kra96,Lun96}, 
where the relaxation of the Si substrate upon hydrogen adsorption was 
represented by one idealized Si phonon coordinate. Although both 
calculations used the same {\it ab initio} energies as a source, the 
results of the dynamical calculations did not agree quantitatively.
However, in both studies the dynamical coupling of the desorption path to 
the Si substrate was low.

In order to help clarifying these various confusing and in fact 
partially conflicting results we have studied the associative desorption 
of D$_2$ from Si\,(100) by determining how the potential energy
at the barrier is distributed over the various degrees of freedom
of this system (hydrogen vibration, rotation, and translational energy,
and vibrations of the Si substrate).
The calculation of PESs is an important prerequisite for
understanding reaction dynamics. For a quantitative analysis, 
however, a calculation of the dynamics is indispensable.
Since the substrate relaxation  plays such an important
role for the hydrogen desorption from Si\,(100), a proper description
of the Si dynamics during the desorption event is required.  
We feel that a reliable dynamical calculation requires the consideration
of the six degrees of freedom of the hydrogen molecule plus at least
two degrees of freedom of the Si surface (dimer stretch and tilt).
A {\em quantum} dynamical treatment of the desorption dynamics taking
these eight degrees of freedom simultaneously into account is presently 
not possible. The record still stands at six dimensions~\cite{Gro95}.
We have therefore performed {\it ab initio} molecular dynamics
calculations to monitor the energy distribution of D$_2$ molecules
desorbing from Si\,(100). This allows us to assess the dynamical 
consequences of the calculated PES. 
We will show that the deuterium molecules leave the surface vibrationally 
hot and rotationally cold, in agreement with thermal desorption
experiments \cite{Kol92}.
Indeed, approximately 0.1~eV of the potential energy at the transition 
state is transferred to substrate vibrations. Still, also the molecules 
receive a noticeable kinetic energy which is at variance with 
a laser-induced desorption experiment \cite{Kol94a}. We will discuss 
possible reasons for this interesting discrepancy.

In the {\it ab initio} molecular dynamics approach \cite{Bock96,DeVit93} 
the forces necessary to integrate the classical equations of motion are
determined by DFT calculations. The exchange-correlation
functional is treated in the generalized gradient approximation 
(GGA) \cite{Per92}. For the hydrogen atoms the full $1/r$ potential is 
used. In previous slab studies the total energies were calculated within
the local density approximation (LDA) with {\it a posteriori} GGA
corrections~\cite{Peh95}. The main effects of using the GGA 
in the complete self-consistent cycle are a small increase of the 
theoretical lattice constant of Si \cite{Mol95} and a slight rise 
in the barrier height from $E_b = 0.3$~eV \cite{Peh95} to
$E_b =0.4$~eV. To correctly represent the up and down buckling of the
clean Si(100) surface we use a (2$\times$2) surface unit cell. 
The Si slab consists of five atomic layers.
The topmost three of them are free to move in the
molecular dynamics simulations, while the remaining two layers are
fixed at their bulk positions. The density-functional calculations
are performed with two {\bf k}-points in the irreducible part of the
Brillouin zone and 40~Ry cutoff energy.
The equations of motion are numerically integrated within a
predictor-corrector scheme with a time step of 1.2~fs.

\begin{figure}[tb]
\begin{center}
\pspicture(0,0)(3,4)
\rput[c](0,2.5){\psfig{figure=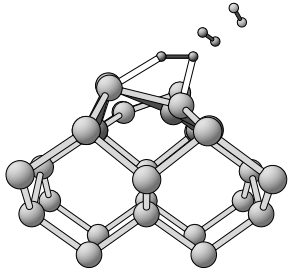,height=4cm}}
\endpspicture
\end{center}
\caption{Snapshots of a trajectory of D$_2$ desorbing from Si\,(100)
starting at the transition state with the Si atoms initially at rest. 
The dark Si atoms correspond to the
Si positions after the desorption event. 
\label{figdes}}
\end{figure}

Since the barrier to associative desorption of hydrogen from Si\,(100) 
is rather high ($E_d \approx 2.5$~eV \cite{Peh95,Sin89,Hoef92}),
there is no sense in performing molecular dynamics calculations starting 
with the deuterium atoms at the adsorption sites because of the extremely
low number of desorption events.
Therefore we started the desorption trajectories
close to the transition state for dissociative adsorption
which was determined in the earlier study \cite{Peh95}.
The desorbing D$_2$ molecule originates from two D atoms which were
bonded to the same Si dimer; this pre-pairing mechanism~\cite{Vit95}
is consistent with the observed first-order kinetics in
experiment \cite{Sin89,Hoef92}. In total we have computed 
42~trajectories of D$_2$ desorbing from Si\,(100). Eight 
trajectories were determined with the Si lattice initially 
at rest, i.e., at a surface temperature of $T_s = 0$~K. Figure~\ref{figdes}
shows snapshots of such a calculated trajectory. The dark Si atoms
correspond to the relaxation of the Si lattice after the desorption event.
Approximately 0.1~eV of the potential energy at the transition state
is transferred to vibrations of the Si lattice which is a rather large
amount compared to what is known from other systems \cite{Dar95}. 
At the transition state the D-D distance is about 40{\%} larger
than the D$_2$ the gas-phase bond length; 
consequently molecular vibrations are excited during the desorption,
as is well known for a long time in associative desorption studies
(see, e.g., ref.~\cite{Dar95}). Due to the strong anisotropy of the
PES, molecular rotations are very effectively quenched during the
desorption event.

The mean kinetic energy of D$_2$ desorbing from Si\,(100) was
determined by laser induced thermal desorption at a rather 
high surface temperature of $T_s \approx 920$~K. 
In order to simulate these experimental conditions,
we have performed {\it ab initio} molecular dynamics calculations
with initial conditions corresponding to the experimental surface
temperature. The system was allowed to equilibrate for more than 500~fs, 
whereby the deuterium atoms were kept close to the transition
state by auxiliary forces $F^c_i$ acting on the deuterium atoms
\begin{equation}
F^c_i(R_i) =  
\left\{ \begin{array}{ccl}
    0 & : & | \Delta  R_i | < \Delta R_i^0 \\
    -k \ \mbox{sign}(\Delta R_i ) \cdot (| \Delta R_i| - \Delta R_i^0) &  
        : & | \Delta R_i| \ge \Delta R_i^0
         \end{array} \right.
\end{equation}\\[-.5cm]
with
\begin{equation}
\Delta R_i \ = \ R_i - R_i^{tr}.
\end{equation}
The force constant $k = 5.2 \times 10^2 \;\mbox{Nm}^{-1}$ 
is taken from the vibrations of the free hydrogen molecule. 
The index $i, \; i = 1,2,3$, denotes the cartesian coordinates, $R_i^{tr}$
are the transition state coordinates of the two D atoms
(see Fig.~\ref{figdes}).
The extra-force-free region is given by
\begin{equation}
\Delta R_i^0 = \left\{ \begin{array}{rcl}
               0.1 \ \mbox{\AA} & : & i = 1,2\\
               0.08 \ \mbox{\AA} & : & i = 3
         \end{array} \right.,
\end{equation}
so that the deuterium atoms in $z$-direction were kept closer to
the transition state than in $x$- and $y$-direction.

The additional potential was switched off when no extra forces were
acting on the D atoms, and the energy distribution of the 
D$_2$ molecules desorbing from the thermal surface was monitored. 
Simultaneously trajectories were run with the additional potential
switched on so that the changes in the trajectories of the Si substrate
atoms due to the desorption event could be followed. This 
trajectory was then also used for the next desorption event.

\begin{figure}[t]
\unitlength1cm
\begin{center}
\pspicture(0,0)(3,10)
\rput[c](.5,8){\psfig{figure=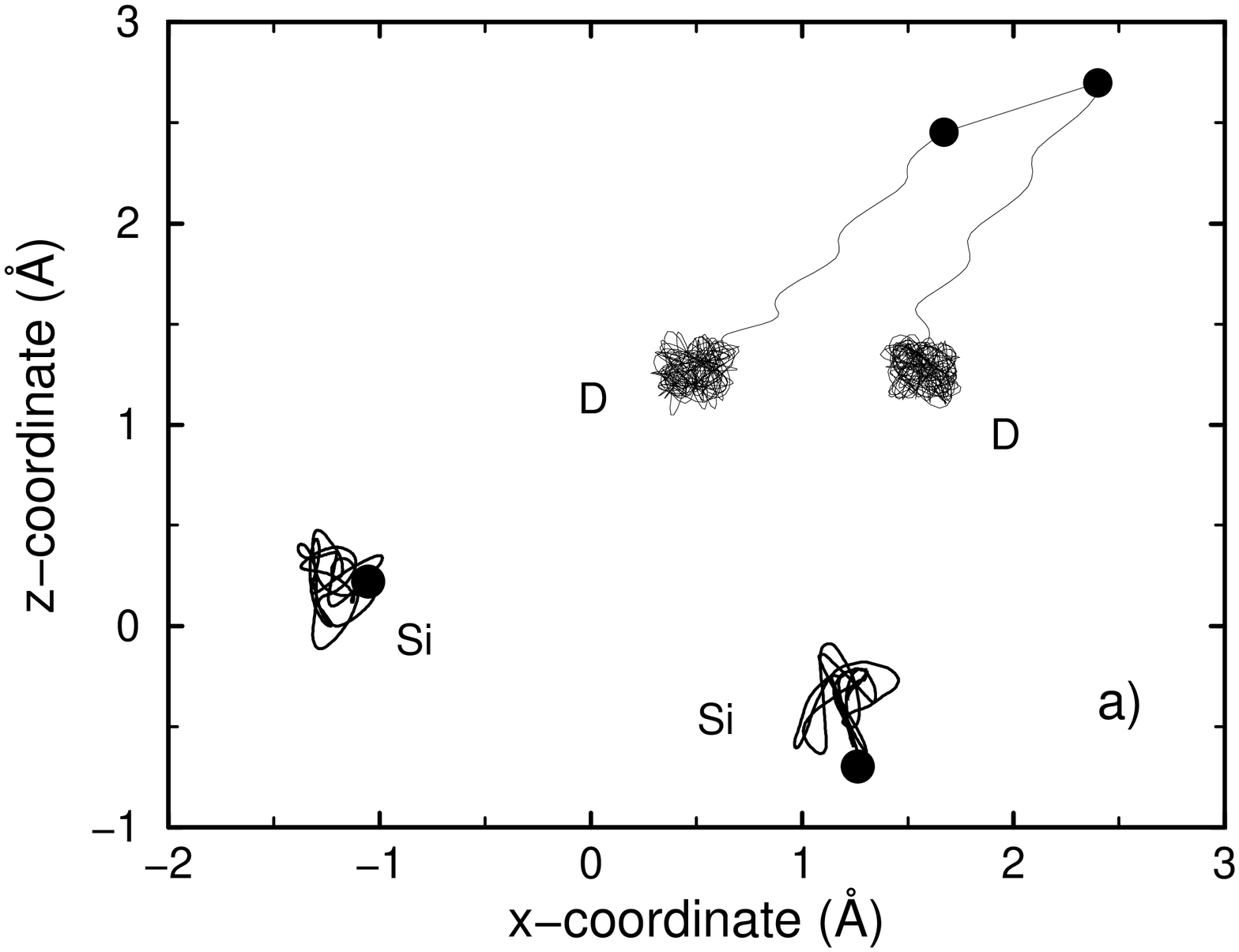,angle=-90,height=5cm}}
\rput[c](.5,3){\psfig{figure=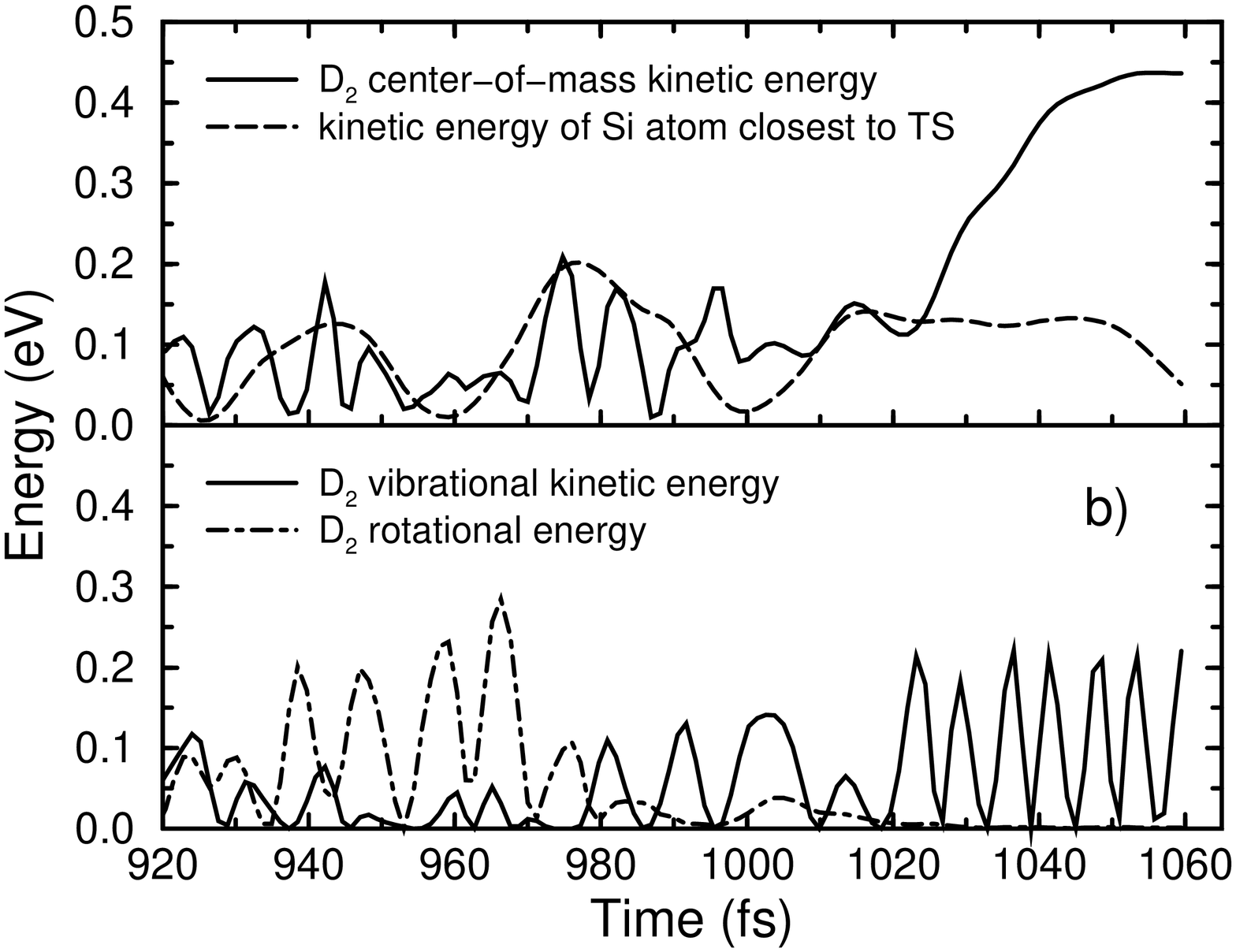,angle=-90,height=5cm}}
\endpspicture

\end{center}
   \caption{Example of a thermal desorption trajectory which was run for
   1060~fs. a) $x$ and $z$
   coordinates of the trajectories of the two deuterium atoms and the
   two Si atoms of the dimer underneath the transition state (TS).
   The final positions of the atoms are marked by the large dots.
   b) energy redistribution as a function of the last 140~fs of
   the run-time of the trajectory. Upper panel,
   solid line: D$_2$ center-of-mass kinetic energy,
   dashed line: kinetic energy of the Si atom closest to the 
   transition state (TS); lower panel,
   solid line: D$_2$ vibrational kinetic energy,
   dash-dotted line: D$_2$ rotational energy.
 }
\label{thermtraj}
\end{figure}

It is true that thermal desorption usually does not correspond
to a genuine equilibrium situation since it is commonly
studied in vacuum. However, desorption events are in general not effected
by the presence or absence of gas molecules in front of the surfaces.
This makes it possible to treat thermal desorption as if it corresponds
to a thermal equilibrium situation \cite{Tul94}. From that it follows 
that we can assume that the phase space density in every point in real-space 
is given by a Boltzmann distribution, in particular at the transition state.
In addition, recrossings of trajectories at the transition state are
very unlikely to occur in the system D$_2$/Si\,(100). 
Due to the high binding energy there are large dissipative effects for 
adsorbing molecules; and for desorbing molecules there are no restoring 
forces in the gas phase. This allows the application of transition
state theory \cite{Cha77} and equivalently also the assumption
of thermal equilibrium for the initial desorption conditions at
the transition state.

In total 34 ``thermal'' desorption trajectories were calculated. 
Figure~\ref{thermtraj} shows an example of such a trajectory with a
total run-time of 1060 fs. For the first 1020~fs the molecular
dynamics was run with the additional potential $V^c$. At $t = 1020$~fs
the extra potential was switched off and the molecule was allowed to
desorb. The projection of the trajectories of the desorbing deuterium 
molecule and of the Si dimer closest to the transition state onto the 
$xz$-plane is shown in Fig.~\ref{thermtraj}a). Clearly the vibrational 
excitation of the desorbing D$_2$ molecule can be seen.
In Fig.~\ref{thermtraj}b) the energy redistribution during
the last 140~fs of the run is plotted. Besides the vibrational excitation
the quenching of the rotational motion can be followed. Also the
acceleration of the D$_2$ center-of-mass is obvious.

\begin{table}[t]
\begin{center}
\begin{tabular}{c|c|c|c}
$\langle E_{tot} \rangle$ & $0.72  \pm 0.17$ eV &
$\langle E_{vib} \rangle$ & $0.11  \pm 0.09$ eV \\ 
$\langle E_{kin} \rangle$ & $0.58  \pm 0.13$ eV & 
$\langle E_{rot} \rangle$ & $0.03  \pm 0.05$ eV \\  
\end{tabular}
\end{center}
\caption{Mean energy distribution averaged over 34 trajectories of D$_2$ 
molecules desorbing from a Si(100) surface at a surface temperature 
of $T_s = 920$~K ($k_B T_s = 0.079$~eV).
\label{tabdes}}
\end{table}

The mean total, kinetic, vibrational, 
and rotational energies of the D$_2$ molecules
averaged over the 34 thermal desorption events are listed in
table~\ref{tabdes}
(note that $k_B T_s = 0.079$~eV at $T_s = 920$~K.). 
The results show vibrational heating, 
i.e. \mbox{$\langle E_{vib} \rangle > k_B T_s$}, and rotational cooling, 
i.e. \mbox{$\langle E_{rot} \rangle < k_B T_s$}, in agreement with the
experiment \cite{Kol92}. The D$_2$ center-of-mass kinetic energy, 
however, is much larger than the experimental value of 
$\langle E_{kin} \rangle^{exp} = 0.165$~eV~$\gtrsim 2k_BT_s$~\cite{Kol94a}. 
The difference between
the experimental and theoretical results corresponds roughly to the
barrier height $E_b = 0.4$~eV. A closer analysis of the trajectories 
reveals that still
approximately 0.1~eV of the potential energy at the transition state is 
transferred to the Si lattice, but due to the Si lattice vibrations
the mean adsorption barrier is increased by roughly the same amount. 
Possible contributions to the discrepancy between theory and experiment
could be:

$(i)$ {\em insufficient statistics} -- Certainly 34 thermal trajectories 
are not enough in order to claim that a true thermal average has 
been performed. However, the widths of our distributions seem
to be converged quite well and we are convinced that by sampling
more points of the phase space at the transition state as initial conditions
the excess translational energy of 0.4~eV would not disappear.

$(ii)$ {\em quantum mechanical effects} -- There are two important quantum
mechanical effects not taken into account by classical molecular dynamics:
tunneling and zero-point effects. 
Low-dimensional quantum dynamical studies of D$_2$/Si(100) have already 
shown that tunneling is strongly suppressed due to the thickness of the 
barrier \cite{Kra96,Lun96}. As for zero-point effects, it has been 
shown~\cite{Kra94} that the sum of
all zero-point energies of the hydrogen molecule at the transition state
approximately equals the zero-point energy of hydrogen in the gas-phase.
In such a situation zero-point effects cancel out as far as the
center-of-mass kinetic energy is concerned since they lead to an almost
constant energetic shift of the minimum energy paths in the hydrogen
coordinates \cite{Gro96}. Hence we infer that quantum mechanical effects 
do not significantly change the kinetic energy of desorbing molecules.

$(iii)$ {\em dissipation channels not considered} -- A further channel for
energy dissipation at surfaces is the excitation of electron-hole pairs.
This channel is not taken into account in the {\it ab initio} molecular 
dynamics simulations where the electrons are assumed to follow the ionic 
motion adiabatically in the electronic ground state.
While electronic excitations of the 
H$_2$ molecule require a too large energy, the excitation
of a surface exciton with a hole in the dangling bond at the
upper Si atom and an electron in the lower Si dangling bond
is in fact a possiblity. This exciton might have 
some resemblence to that at the (110) surface of GasAs~\cite{pankratov}
The described surface exciton at Si\,(100) will strongly couple to the
lattice and bring the Si dimer in a geometry parallel to the surface.
Thus, the H$_2$ desorption ends with an electronic state and geometry
0.2 eV  higher than the ground state, and this energy
would be taken out of the H$_2$ kinetic energy.

$(iv)$ {\em limitations of the GGA functional} -- It was recently shown 
that the GGA-PW91 functional used in our study underestimates the H$_2$ 
elimination barriers from silanes, i.e., small molecules \cite{Nac96}. 
On the other hand, energies for surface systems calculated from cluster 
models are sometimes seriously in error (see, e.g., Ref.~\cite{Whi96}) 
and typically converge very slowly with cluster size.
Hence results for small molecules are not directly transferable to
surface systems.
Still the validity of the GGA-PW91 functional for the barrier heights of
H$_2$/Si(100) certainly remains an open question since the application
of GGA functionals for dissociation at surfaces is still a rather new 
field.

$(v)$ {\em experimental uncertainties} -- The desorption experiments were 
done with a Si sample with a high dopant concentration 
of $n \approx 10^{19}$~cm$^{-3}$. Therefore the Si\,(100) surface 
used in the experiment might have been metallic.
It is well possible that for such a system the geometry of the clean surface
is different from that of an undoped substrate.
Thus, the final state of the desorption is changed and possibly also
the geometry and energy of the transition state. 
In fact, also in Ref.~\cite{Bra96a} it had been speculated that
high doping might influence the barrier heights
in order to explain that the sticking probability for hydrogen molecules
at Si\,(100)  is much higher for a highly doped sample~\cite{Kol94b}
than for a weakly  doped system~\cite{Bra96a}.

In addition, recent {\it ab initio} calculations indicate
that laser melting of bulk silicon leads to a transition to a
metallic liquid state~\cite{Sil96} which could also occur during 
the laser-induced desorption used in the experiment. This would also 
change the geometrical and electronic structure at the transition state.
Furthermore, at a metallic surface the excitation of electron-hole
pairs might affect the desorption dynamics and, for example,
enhance the probability of the above noted exciton state.

In conclusion, we have presented an {\it ab initio} molecular dynamics
study of the desorption of D$_2$ from Si\,(100). The calculations 
give a qualified description of the desorption of hydrogen
from undoped and good quality Si\,(100). While the energy 
distribution of the desorbing molecules in the vibrational and rotational
degrees of freedom is in agreement with experiment~\cite{Kol92}, the excess
translational energy is too large compared to the experiment of
Ref.~\cite{Kol94b}. This discrepancy might be caused by uncertainties in 
the determination of the interaction potential; however, it could also be 
due to the high doping of the Si sample used in the experiment of
Ref.~\cite{Kol94b}.

We gratefully acknowledge a grant of computer time at the
Cray~T3D of the Konrad-Zuse-Zentrum, Berlin. We thank M. Fuchs for
providing us with a GGA-pseudopotential of Si and E. Pehlke, P. Kratzer, 
W. Brenig, and E. Hasselbrink for helpful discussions.

\end{document}